\documentclass[twocolumn]{IEEEtran}

\usepackage{graphicx}
\usepackage{subfigure}
\usepackage{threeparttable}
\usepackage{float}
\usepackage{epsfig}
\usepackage{epstopdf}
\usepackage{amsmath}
\usepackage[ruled]{algorithm2e}
\usepackage{xcolor}
\usepackage{flushend}

\newcommand{\fq}[1]{\textcolor{black}{#1}}

\begin{document}

\title{\Large{Analytical Modeling of UAV-to-Vehicle Propagation Channels in Built-Up Areas}}

\author{Zhuangzhuang~Cui,~Ke~Guan,~C\'esar~Briso,~Danping~He,~Jianqiao~Cheng,~Zhangdui~Zhong,~and~Fran\c cois~Quitin
        \thanks{Manuscript received xx xx, 2019; revised xx xx, 2019; accepted xx xx, 2019. This work was supported by the Belgian Fonds de la Recherche scientifique-FNRS (FRS-FNRS) under grant R.M008.18 and the National Key R\&D Program of China under grant No. 2016YFB1200102-04. \emph{(Corresponding author: Ke Guan.)}}
        \thanks{Z. Cui, K. Guan, D. He and Z. Zhong are with State Key Lab of Rail Traffic Control and Safety, Beijing Jiaotong University, Beijing 100044 China (e-mail: cuizhuangzhuang@bjtu.edu.cn, kguan@bjtu.edu.cn).}
        \thanks{C. Briso is with ETSIS Telecommunications, Universidad Polit\'ecnica de Madrid, 28031 Madrid, Spain (e-mail: cesar.briso@upm.es).}
        \thanks{J. Chen and F. Quitin are with Brussels School of Engineering, Universit\'e Libre de Bruxelles, Brussels, Belgium (e-mail: Jianqiao.Chen@ulb.ac.be).}}

\markboth{IEEE WIRELESS COMMUNICATIONS LETTERS,~Vol.~xx, No.~xx, xx~2019}%
{Z.Z. Cui \MakeLowercase{\textit{et al.}}:Analytical Modeling of UAV-to-Vehicle Propagation Channels in Built-Up Areas}

\maketitle

\begin{abstract}
\fq{This letter presents an analytical path loss model for air-ground (AG) propagation between unmanned aerial vehicles (UAVs) and ground-based vehicles. We consider built-up areas, such as the ones defined by ITU-R. }The three-dimensional (3D) path loss model \fq{is based on propagation} conditions and essential parameters are derived by \fq{using geometric methods}. Owing to the generality, the analytical model is capable of arbitrary deployments of buildings\fq{,} such as suburban, urban and dense urban. \fq{The analytical model is evaluated numerically, }and validations conducted by ray-tracing simulations show the high accuracy of the proposed model. The closed-form analytical formulas provide a useful tool \fq{for quick and accurate prediction of UAV-to-vehicle} propagation channels.
\end{abstract}


\IEEEpeerreviewmaketitle

\section{Introduction}
\IEEEPARstart{T}{he} advent of the internet of everything (IoE) in the fifth generation and beyond (5GB) wireless systems will give rise to a new category of use cases termed \fq{air-ground} (AG) integrated communication, where aerial and terrestrial terminals are interconnected. As an example, unmanned aerial vehicles (UAVs) serving as users can connect \fq{to a }cellular base station (BS) \cite{b2}, \fq{but can also} act as aerial BSs to provide data services for ground terminals \fq{(such as smartphones or vehicles)}. The \fq{UAV-to-vehicle} communication is \fq{anticipated to play a critical role} in the traffic control system and emergency communication. UAVs can assist vehicular networks by acting as intermediate relays or BSs. UAVs can also be deployed in an area where a disaster occurs, and therefore acts as bundle carriers and relays. However, the requirements in \fq{control}, communication and navigation pose \fq{challenging issues to system designers of integrated networks. In particular, modeling the UAV-to-vehicle propagation channel remains an open issue, especially in built-up areas. Such environments have} many potential scatterers composed of buildings, \fq{which will cause the channel to }experience severe multipath effects. \fq{Additionally, because the altitude of the UAV may change, it is important to consider three-dimensional channel models that take the altitude of the transceiver into account.}

The literature on AG channel characterization and modeling is \fq{sparse}, \fq{and the existing work focuses on measurement-based} and geometry-based approaches. Measurement campaigns were conducted for different scenarios such as urban, over-water and mountain \fq{area }in \cite{b3}\fq{, where tapped} delay line (TDL) models were presented. \fq{However, since the measurement were carried out at a single altitude, they are insufficient for setting up an altitude-dependent model. }\fq{In \cite{b4, b5}, authors discuss a} more generic geometry-based stochastic model (GBSM)\fq{,} with the assumption that scatterers \fq{are distributed stochastically on a} cylinder and \fq{an} ellipsoid, respectively. In the above methods, \fq{measurement based on the specific scenarios lead} to a lack of generality. \fq{Moreover, GBSM have trouble reflecting realistic performance metrics that can be observed in real scenarios. }As a tradeoff, \fq{three-dimensional analytical model become} more promising for quick and accurate prediction \fq{of UAV-to-vehicle channels}.

\fq{Despite their obvious importance, UAV-to-vehicle channels have received little} attention. Authors in \cite{b6} focused \fq{on transmission rates and packet losses} of UAV-to-car communication\fq{s} while \fq{ignoring} the complex channel characteristics. \fq{In this paper, we present the importance of AG channel modeling and propose an analytical channel model for UAV-to-vehicle communication. }The main contributions of this work can be summarized \fq{as follows}: (1) a novel path loss model \fq{is} proposed with consideration of the scenario characteristics defined by ITU-R, (2) the critical parameters of the \fq{multipath channel are} derived, and (3) validations by ray-tracing simulations \fq{are} conducted in the reconstructed Manhattan scenario based on \fq{a} real environment \cite{b7}.

The remainder of this letter is organized as follows. In Section II, we introduce the general scenarios defined by ITU-R and present \fq{the derivation of} the analytical model. Section III \fq{presents} numerical results and analysis. Besides, validations based on ray-tracing simulations \fq{of} the proposed model are conducted. Conclusions are drawn in Section IV.
\begin{figure}[htbp]
\centering
\subfigure[]{\includegraphics[width=1.9in]{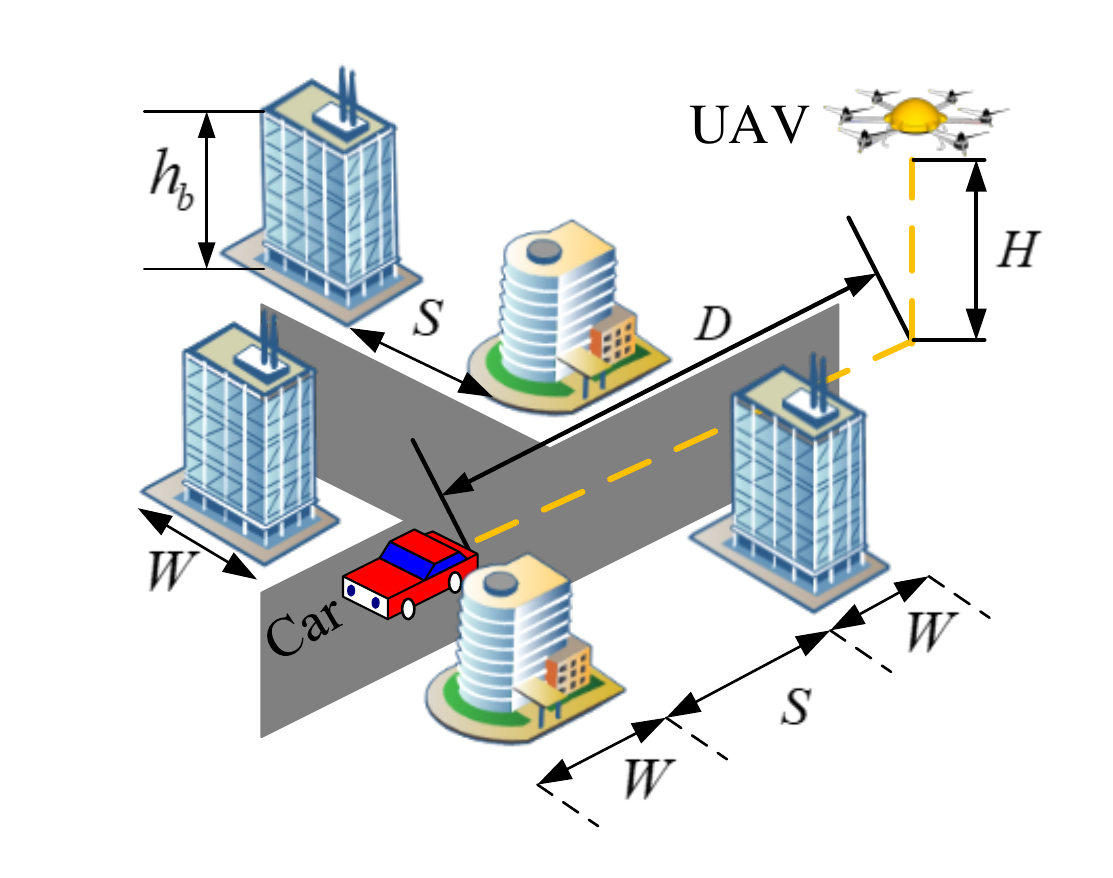}}
\subfigure[]{\includegraphics[width=1.7in]{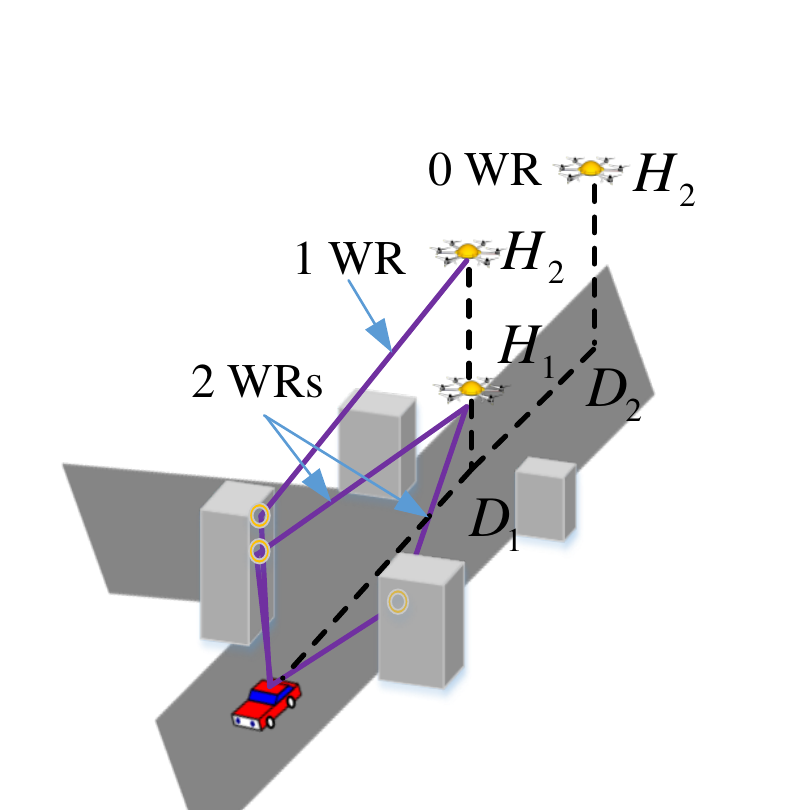}}
\caption{(a) \fq{Representation of the} built-up scenario with square buildings (side length-$W$) and streets (width-$S$) where the altitude of \fq{the} UAV is $H$; the horizontal distance between \fq{the} UAV and \fq{the vehicle} is $D$; the height of building is $h_b$; (b) Propagation conditions with the focus \fq{on} the wall reflection. }
\label{fig_sim}
\end{figure}
\section{Propagation Modeling in Built-Up Areas}
\subsection{Scenario Description}
In order to set up a generic path loss model in BU areas, the statistical ITU-R Rec. P.1410 model for building deployments is adopted \cite{b8}. An advantage of this model is that the area can be modeled without precise information concerning building shapes and distribution\fq{s}. The statistical model requires only three empirical parameters describing the built-up area: the ratio of the land area covered by buildings to the total land area ($\alpha$), the average number of buildings per unit area ($\beta$), and a parameter for determining the distribution of building height ($\gamma$). \fq{Typical} values \fq{for these} three empirical parameters are listed in Table I. The probability density function (PDF) for the building height \fq{is based on a} Rayleigh distribution \fq{and is} given by
\begin{equation}
p_r(h_b) =\frac{h_b}{\gamma}\exp(-\frac{h_b^2}{2\gamma^2}).
\end{equation}

In this study, \fq{the virtual-city environment is chosen similar to a }Manhattan grid \cite{b9}. \fq{Fig. 1(a) shows such an environment, composed of }an array of structures \fq{of width $W$ and inter-building spacing $S$.} \fq{The parameters $W$ and $S$ are measured in meters} and can be linked to the ITU-R statistical parameters, since \fq{by definition} $\alpha$ = $(N_bW^2)$/$(1000Q)^2$ \fq{(where $Q$ is the map side measured in kilometers)}, and $N_b$ is the number of \fq{buildings}. \fq{The parameters $N_b$ can be linked to the ITU-R parameters by} $\beta=N_b/Q^2$. Thus, we can calculate $W$ and $S$ as \fq{follows}: $W=1000\sqrt{\alpha/\beta}$ and $S=1000/\sqrt{\beta}-W$. Fig. 1(a) shows the \fq{environment representation with essential parameters for} UAV-to-vehicle communication in built-up areas. Note that parameters in Table I are the typical values of standard cities. In channel modeling, we can use special values which are obtained according to the actual sizes of environments.
\begin{table}[!htbp]
  \centering
  \caption{Typical Parameters of Standard City from ITU-R P.1410 \cite{b8}} \label{table1}
\begin{tabular}{|c|c|c|c|}
  \hline
\textbf{Environment} & \textbf{$\alpha$} & \textbf{$\beta$} & \textbf{$\gamma$}\\
    \hline
Suburban & 0.1  & 750 & 8 \\
\hline
Urban & 0.3 & 500 & 15\\
    \hline
Dense Urban & 0.5 & 300 & 20 \\
\hline
High-rise Urban  & 0.5  & 300 & 50 \\
  \hline
\end{tabular}
\end{table}

\subsection{UAV-to-Vehicle Path Loss Model in Built-Up Scenarios}
\fq{For the BU case}, we found that the channel is well modeled by \fq{a} two-ray (TR) model consisting of the direct path and the ground reflection\fq{,} plus intermittent multipath components (MPCs) which are determined according to the altitude of UAV and the horizontal distance. Note that due to the long path travelling, the power of diffractions and high-order reflections is insignificant, which can be neglected in the modeling of path loss. As shown in Fig. 1(b), at a distance of $D_1$ and \fq{a} lower altitude $H_1$, the number of wall reflection (WR) may be \fq{two}. When the height increases to $H_2$ (at the same distance $D_1$), the reflection may become \fq{only one} since the height of potential reflected point increases as the height of \fq{UAV increases. Another geometrical consideration for the WR} the space between buildings, for example, at a distance of $D_2$. \fq{At this location}, there is no WR regardless of the altitude of UAV. Thus, according to the different conditions of propagation and the TR model in \cite{b10}, the path loss model for UAV-to-vehicle propagation channels can be expressed as
\begin{equation}
\label{PL_UAVV}
\begin{aligned}
PL_{BU} [dB] = & 20\log_{10}(\frac{4\pi d}{\lambda}|\underbrace{1+\Gamma_g \exp(i \Delta \varphi_g)}_{\rm {TR}} + \\& \underbrace{\Gamma_b\sum^{}_{Num=0,1,2} \exp(i \Delta \varphi_b)}_{\rm {WR}}|^{-1})
\end{aligned}
\end{equation}
where $\lambda$ is the wavelength and $d$ is the link distance. $\Gamma_b$ and $\Gamma_g$ are reflection coefficient for the building and \fq{the} ground, respectively. $\Delta\varphi_{b}$ and $\Delta\varphi_{g}$ are the phase differences between the wall or ground reflection and the line-of-sight (LOS), respectively, which can be calculated by
\begin{equation}
\label{delta_fi}
 \Delta\varphi_{b/g}=\frac{2\pi}{\lambda}(d_{LOS}-d_{Ref.,b/g})
\end{equation}
where $d_{LOS}=\sqrt{D^2+(H-h_v)^2}$ where $h_v$ is the antenna height mounted on the top of vehicle. $d_{Ref.,b}$ and $d_{Ref.,g}$ are the path length of building reflection and ground reflection, respectively. According to the relationship of geometry, $d_{REF,b}$ can be calculated by
\begin{equation}
d_{Ref.,g}=\sqrt{D^2+(H+h_v)^2}
\end{equation}
In this paper, we assume the ground vehicle is at the center of the road while the UAV \fq{also flies along the center of the road}, which means \fq{that} the distance{s to} the buildings on either side \fq{of the vehicle (and their ground projections)} \fq{are} $S/2$. Thus, the path length of two-side wall reflections is \fq{the }same and can be calculated by
\begin{equation}
d_{Ref.,b}=\sqrt{S^2+D^2+(H-h_v)^2}
\end{equation}
where the calculation is derived based on the geometry.
\begin{figure}[htbp]
  \centering
 \subfigure[]{\includegraphics[width=1.7in]{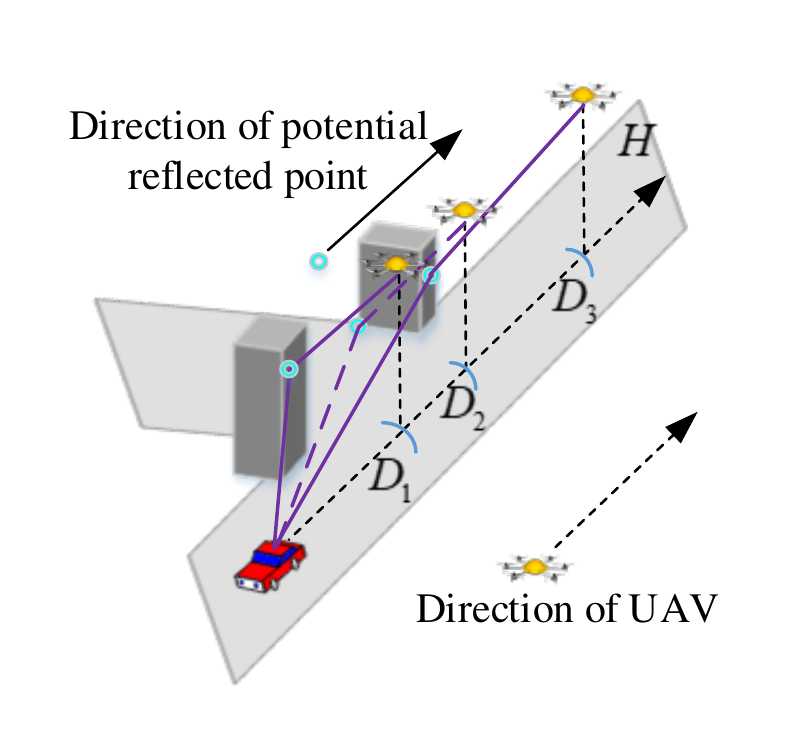}}
\subfigure[]{\includegraphics[width=1.7in]{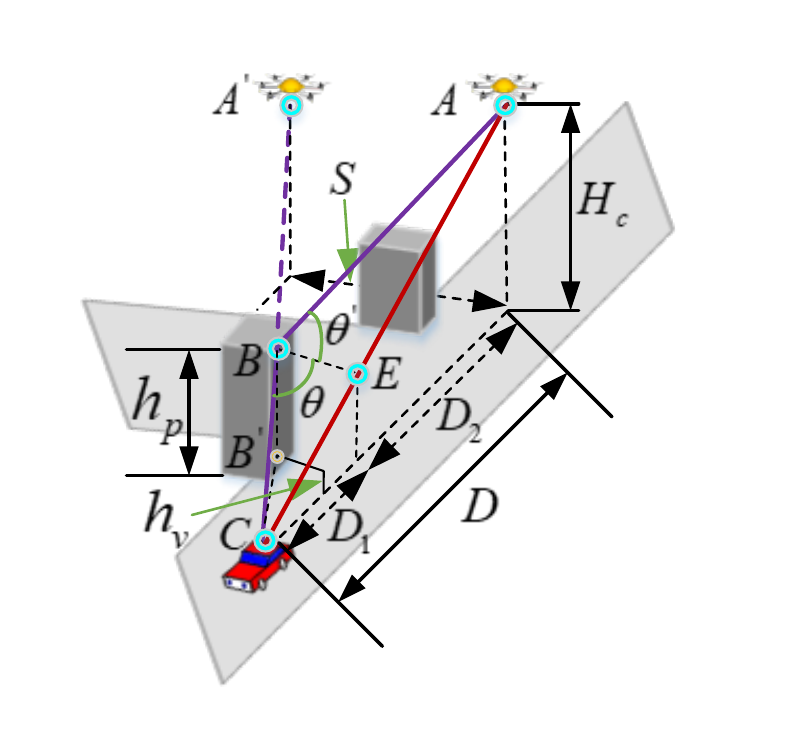}}
  \caption{(a) Potential reflected point moves as the horizontal distance changes; (b) Determining the critical altitude of UAV using \fq{geometrical} relationship\fq{s}.}\label{critical_proof}
\end{figure}

\subsection{Periodic Characteristic of the Wall Reflection}
As shown in Fig. 2(a), \fq{when the UAV flies along the road the reflection point of the WR changes. If the UAV and the ground vehicle are static, the WR can be determined according to the geometry of the environment. If the UAV (or the vehicle) moves, the WR behavior will become periodic as the UAV (or the vehicle) travels from building to building. As can be seen in Fig. 2(a), there is a WR when the potential reflection point is on the wall. When the potential reflection point moves to the space in between buildings, there is no WR. }Note that \fq{we assume here that} the altitude of UAV is lower than the critical altitude which is derived in the next subsection. In addition, \fq{it should be noted that the WRs happen from the buildings on both sides of the road. }

\subsection{Determining the Critical Altitude of UAV}
As shown in Fig. 1(b), \fq{increasing} the altitude of UAV will lead \fq{the WR to disappear}. Thus, it is \fq{important} to calculate the critical altitude of UAV. In other words, we want to determine whether a WR occurs based on the heights of the buildings, vehicle and UAV. Note that the \fq{following} deduction assumes the potential \fq{reflection} points are on the wall. We introduce the heights of the \fq{buildings on both sides of the road as} $h_{p}^1$ and $h_{p}^2$, respectively. For the UAV-to-vehicle case, the height of the vehicle is fixed and lower than the building \fq{height} so that the vital issue to determine the number of WR is to find the critical altitude of the UAV. \fq{We show that the critical altitude of the UAV is given by}
\begin{equation}
H_c= 2h_{p}-h_v
\end{equation}
where $H_c$ is the critical altitude of UAV, $h_v$ is the antenna height of vehicle and $h_p$ is the height of building producing the potential reflections. Thus, when the altitude of UAV $H\le 2h_{p}^1-h_v$ and $H\le 2h_{p}^2-h_v$, there will be \fq{two} WRs. Similarly, when $H\le 2h_{p}^1-h_v, H\ge 2h_{p}^2-h_v$ or $H\ge 2h_{p}^1-h_v, H\le 2h_{p}^2-h_v$, there will \fq{only be one} WR. \fq{When} $H\ge 2h_{p}^1-h_v$ and $H\ge 2h_{p}^2-h_v$\fq{, there is no WR. }

\emph{Proof:} As shown in Fig. 2(b), the mirror symmetry point of \fq{the} UAV about the wall is $A'$. \fq{Eq. (5) can easily be deduced by simple geometry. }In order to obtain the critical altitude of UAV, we assume the reflected point is on the top edge of the building with a height of $h_p$. According to Snell's law, $\theta$=$\theta'$. We make a vertical line from point $B$ and intersect \fq{it} with \fq{the LOS at point} $E$. \fq{Since} $\theta$=$\theta'$, $BE \bot AC$ and $BE=BE$, \fq{we can state that} $\triangle ABE = \triangle CBE$. \fq{Therefore it can be established that $AB=CB$ and $D_1=D_2=D/2$. }Thus, $BC$ can be calculated by
\begin{equation}
BC=\sqrt{(D/2)^2+(S/2)^2+(h_p-h_v)^2}
\end{equation}
where $BC$ is actually equal to $d_{Ref.,b}/2$. By substituting Eq. (7) into Eq. (5), we can obtain $H_c=2h_p-h_v$.


\section{Numerical Results and Validation }
\subsection{Numerical Results}
\fq{This section provides simulation results. } We set the \fq{carrier} frequency at 4 GHz which is in the typical range for UAV and vehicle communications \cite{b3,b6}. For simplification, we assume $\Gamma_b$ = $\Gamma_g$ = 1\fq{. The} scenario is urban, and the parameters are listed in Table I so that we can calculate $W=24.5$~m and $S=20.2$~m. The altitude of UAV is set \fq{to} 50~m, and the height of the vehicle is 1.5~m. Fig. 3 shows the periodicity of WR because of the changing horizontal \fq{position of the UAV}. When the number of multipath is 2, the WR \fq{are} non-existent. Note that the height\fq{s} of the building\fq{s are} randomly generated and the critical altitudes are calculated \fq{with} Eq. (6). Thus, the WR can be 0, 1 and 2 because of the random heights of the buildings.
\begin{figure}[htbp]
  \centering
  \includegraphics[width=2.6in]{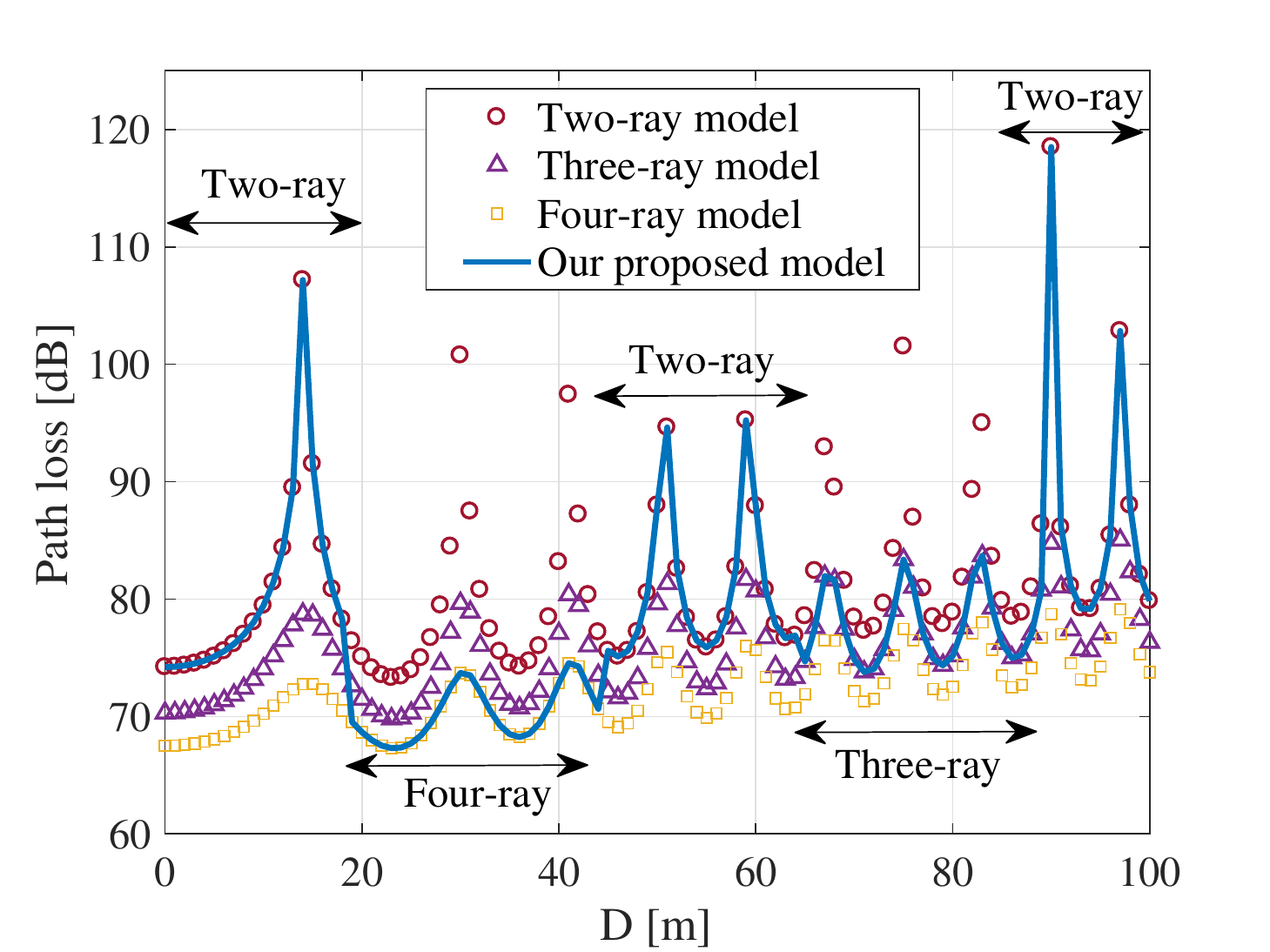}\\
  \caption{Path loss result in urban scenario when $H=50$ m and $f=4$ GHz.}
  \label{55}
\end{figure}

Fig. 4 shows that as the altitude of UAV rises, the \fq{number of multipath changes linearly}, which is different from the periodic characteristics when the horizontal \fq{position of the UAV changes}. In the simulation, the two critical altitudes \fq{(corresponding to the buildings on the two sides of the road)} are calculated so that when the altitude of UAV is lower than the smaller critical altitude, the number of WR is 2; when the UAV altitude is between \fq{the two critical altitudes}, the number of WR is 1. Once the altitude higher than \fq{largest} critical altitude, there is no WR.
\begin{figure}[htbp]
  \centering
  \includegraphics[width=2.6in]{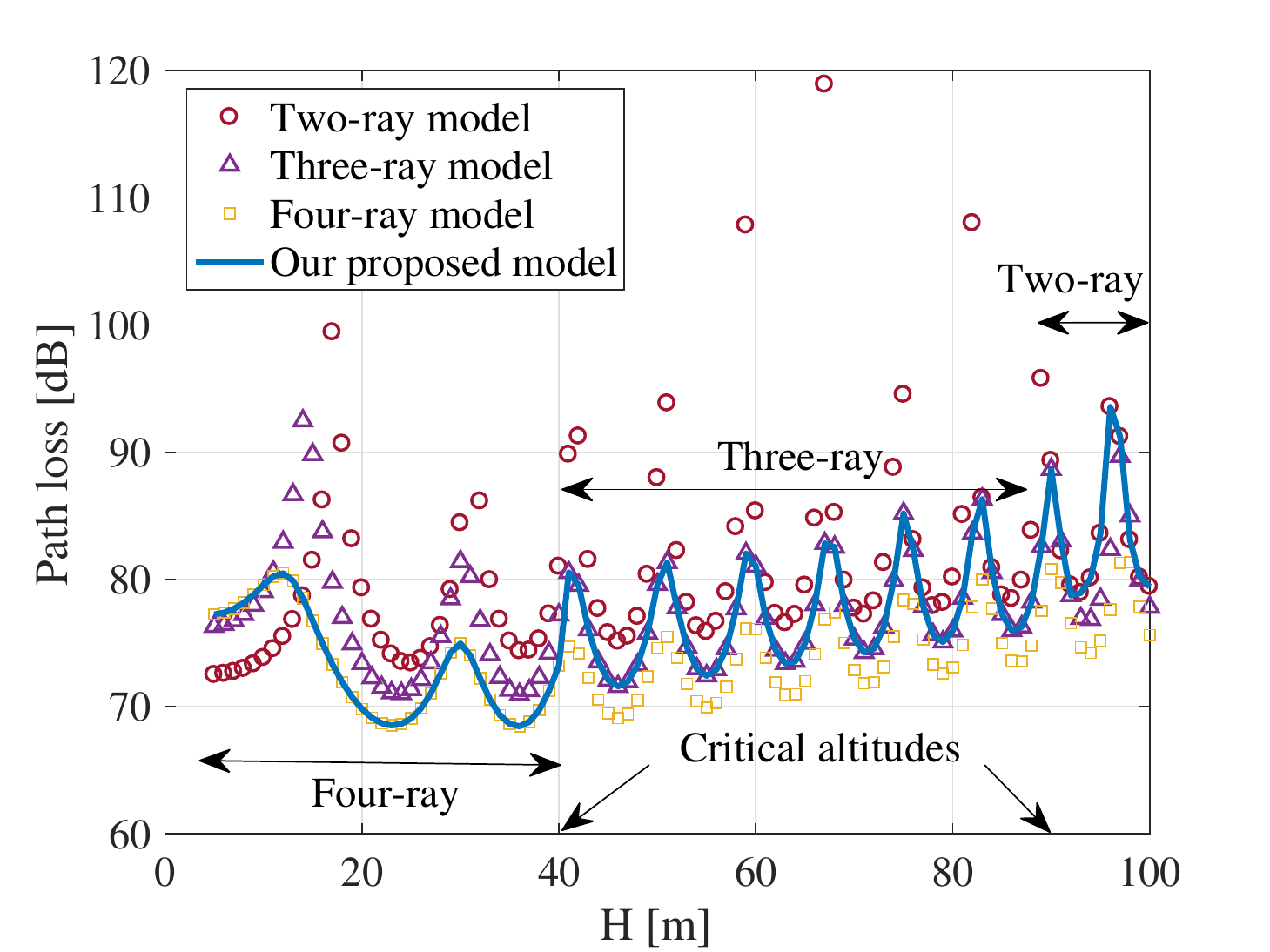}\\
  \caption{Path loss result in urban scenario when $D=50$ m and $f=4$ GHz.}
  \label{55}
\end{figure}

For different scenarios, we compare the path loss \fq{in} Fig. 5. We obtain the path loss result when $H=$~50~m and $D$ \fq{varies} from 0 to 100 m. The sizes of \fq{the }scenarios are obtained according to the parameters in Table I. The cumulative probability density functions (CDFs) are plotted and optimally fitted by Normal distribution. We observe that the mean values ($\mu$) of them are similar \fq{and} around 72.5-75 dB. However, the standard deviation $\sigma$ presents a noticeable difference. The trend is\fq{ the following:} as the density of building decreases, that is, from dense urban to urban to suburban, $\sigma$ changes from 6.15 dB to 8.03 dB to 9.48 dB. It is hard to understand since we take it for granted that the density of building decreases, the level of path loss fluctuation (the physical meaning of $\sigma$) will become smaller owing to fewer rays. In fact, due to the larger space between buildings in the lower density scenario, the periodic characteristic of multipath is more evident for the UAV-to-vehicle channel so that fluctuations become larger. Another reason is that more WRs in the higher density scene\fq{s} lead to smaller fluctuations of the path loss.
\begin{figure}[htbp]
  \centering
  \includegraphics[width=2.6in]{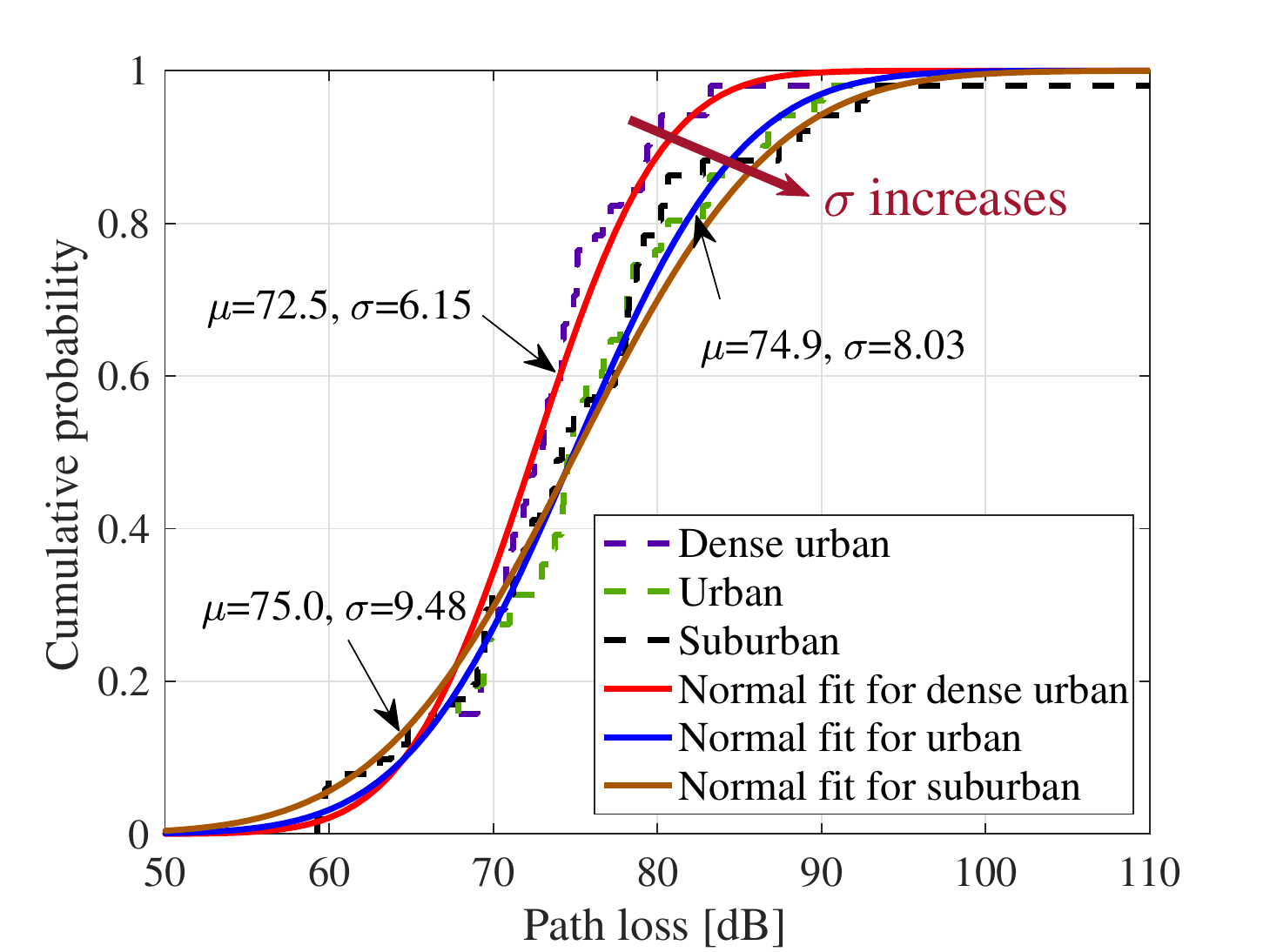}\\
  \caption{CDF of path loss for different scenarios when $H$=50 m, $D$=0-100 m.}
  \label{55}
\end{figure}
\subsection{Validation for Proposed Model by Ray Tracing}
We validate our proposed model by \fq{ray-tracing simulation, }which \fq{has been verified as a method for channel modeling }by massive measurements in Europe, South Korea, and America. The digital map shown in Fig. 6(a) is based on \fq{a} real environment in Manhattan, New York and obtained from \fq{the} \emph{OpenStreet} website. The CDF of \fq{the} building height is plotted in Fig. 6(b). The result ($\gamma$) of Rayleigh fit is 87.3. Besides, the size of the scenario is marked. The simulation is conducted at 4 GHz, the altitude of UAV and vehicle is \fq{set to} 200~m and 1.5~m, respectively, and the horizontal \fq{location of the UAV goes} from 0 to 225~m.
\begin{figure}[htbp]
  \centering
   \subfigure[]{\includegraphics[width=1.5in]{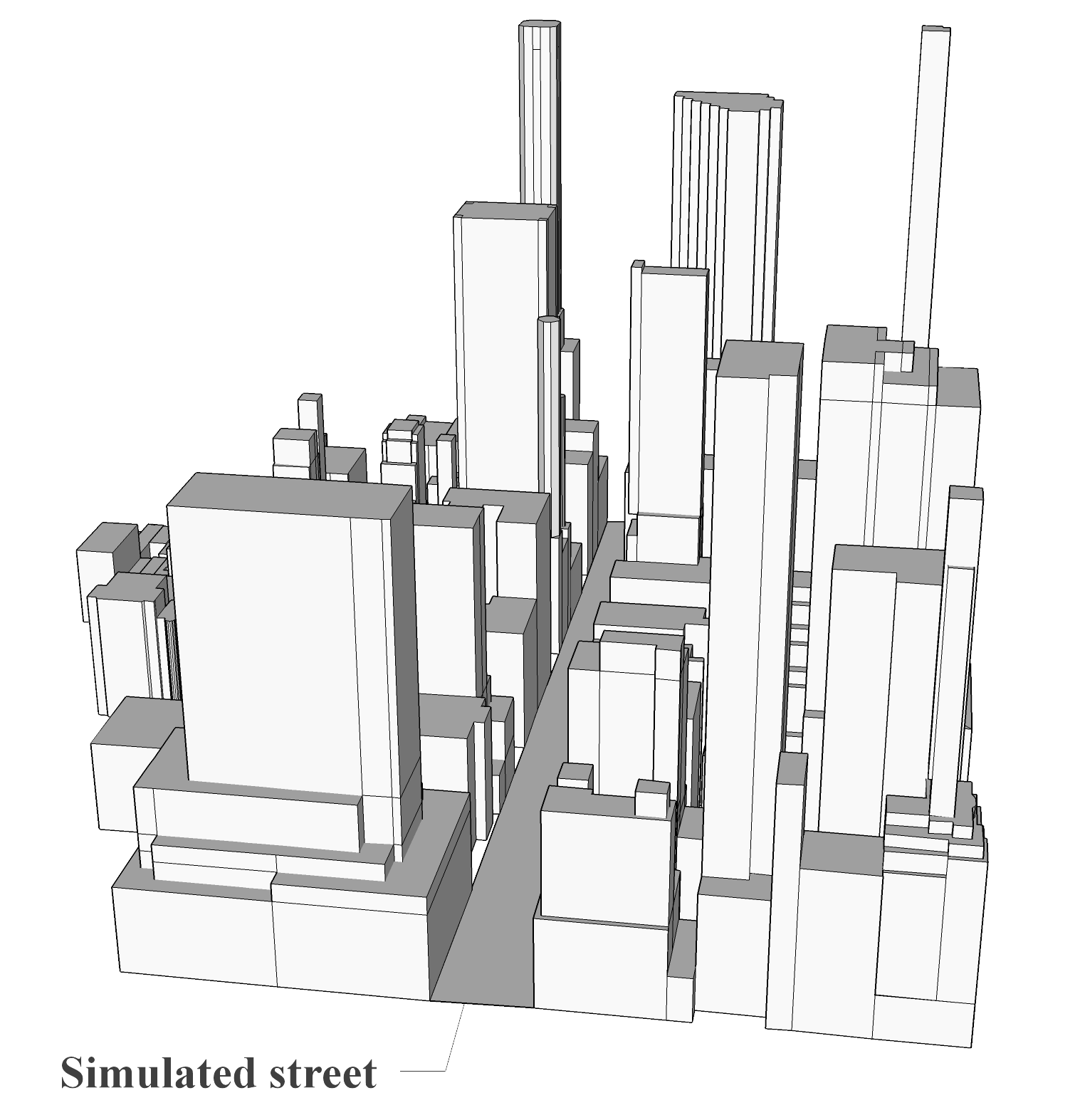}}
   \subfigure[]{\includegraphics[width=1.8in]{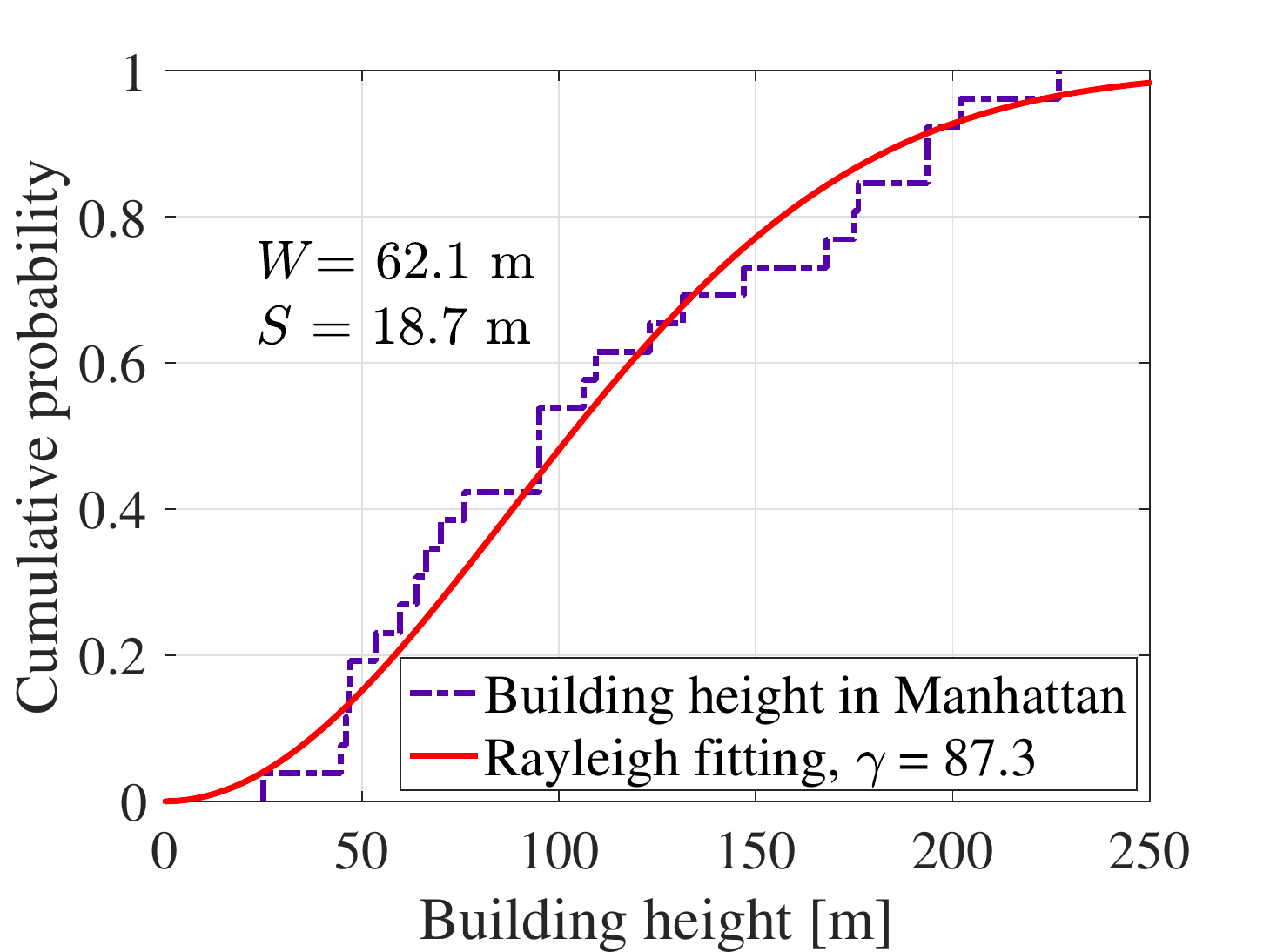}}\\
  \caption{(a) SketchUp model for Manhattan; (b) CDF of the building height.}
  \label{55}
\end{figure}
\begin{figure}[htbp]
  \centering
   \subfigure[]{\includegraphics[width=1.72in]{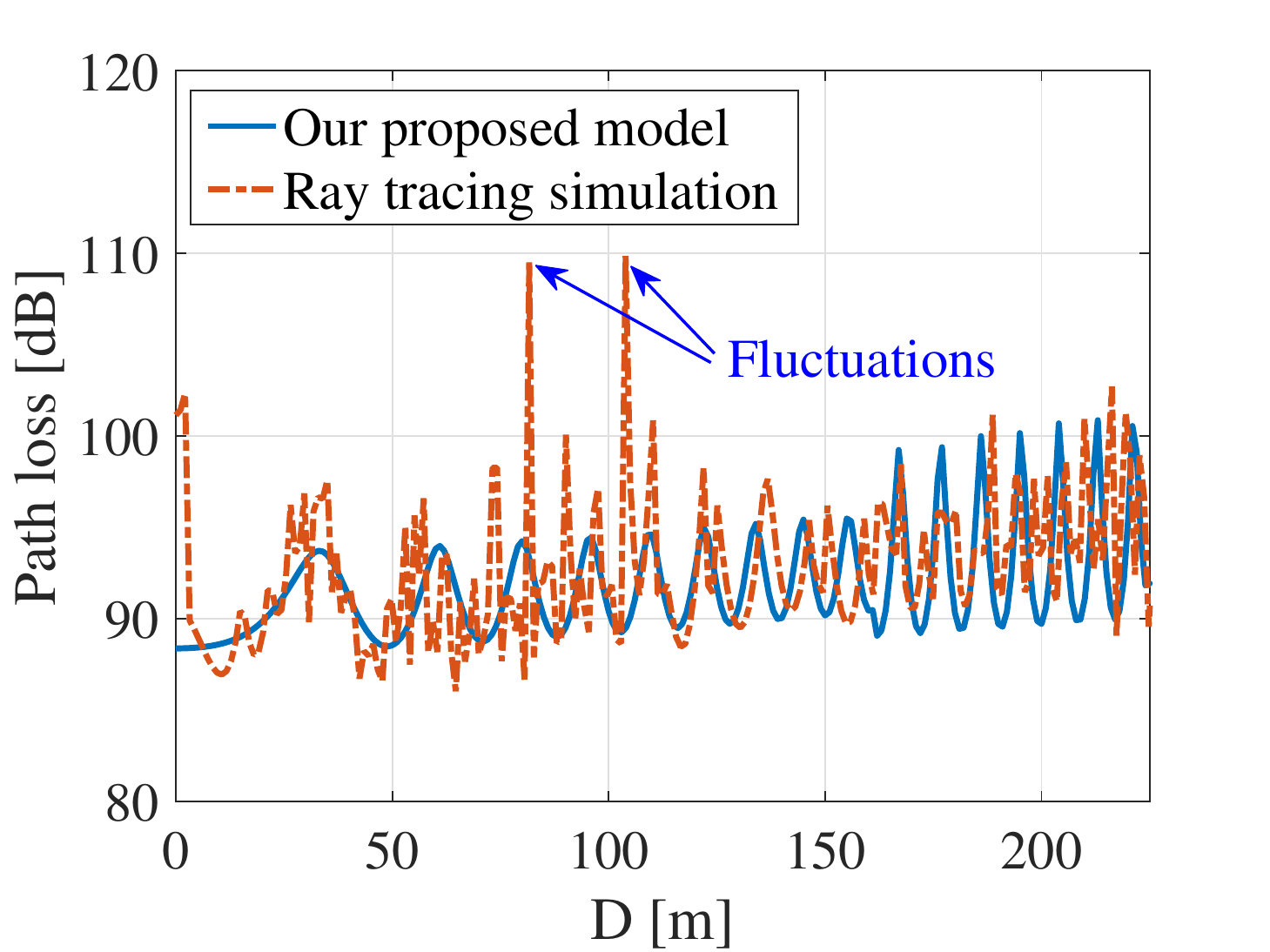}}
   \subfigure[]{\includegraphics[width=1.72in]{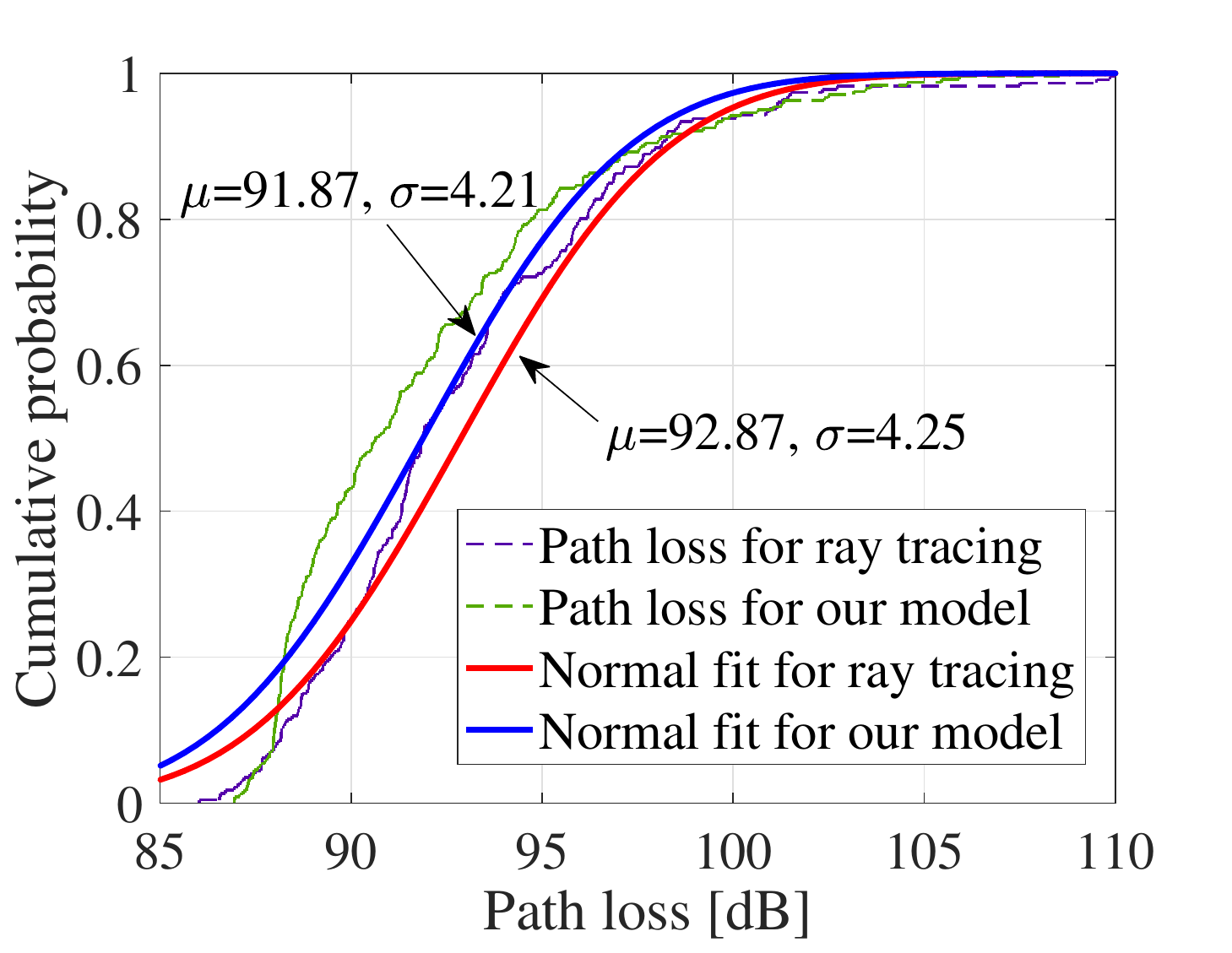}}\\
  \caption{(a) Path loss for our model and ray tracing; (b) CDF of path loss.}
  \label{55}
\end{figure}

\fq{Fig. 7 shows a comparison between the proposed model and the ray-tracing simulations}. The results \fq{indicate} that our proposed model \fq{shows} good agreement with the ray-tracing result. The mean values and standard deviations of \fq{the} path loss for our model and ray tracing is 91.87 dB, 4.21 dB, and 92.87 dB, 4.25 dB respectively. Similar results show the accuracy of our proposed model. However, due to the randomness of building height in our model and the relative irregularity of Manhattan scenario, the large fluctuations \fq{of the path loss at} two locations \fq{are not well taken into account with our model}. In general, \fq{however}, our model can provide a \fq{reasonably} accurate prediction \fq{for UAV-to-vehicle} communication.

\section{Conclusion}
In the paper, we \fq{proposed} an analytical path loss model for predicting the propagation channel of UAV-to-vehicle scenarios. The corresponding parameters are derived \fq{based on a geometrical} method. We found that the number of multipath changes by the way of periodicity and linearity in the horizontal and vertical dimension, respectively. Besides, the standard deviation of path loss decreases as the density of building increases, which is different from traditional channels. Finally, the accuracy of the proposed path loss model is validated by ray-tracing simulations conducted in \fq{a realistic }environment. The study of this paper can be used to \fq{guide }the design and deployment for UAV-to-vehicle communications.

\bibliographystyle{IEEEtran}

\end{document}